\documentclass[useAMS,usenatbib]{mn2e}
\usepackage{graphicx}
\usepackage{amsmath}
\usepackage{epstopdf}
\usepackage{amssymb}
\usepackage{extarrows}
\usepackage{color}
\usepackage{bm}

\def\be{\begin{equation}}
\def\ee{\end{equation}}
\def\ba{\begin{eqnarray}}
\def\ea{\end{eqnarray}}
\def\go{\mathrel{\raise.3ex\hbox{$>$}\mkern-14mu
             \lower0.6ex\hbox{$\sim$}}}
\def\lo{\mathrel{\raise.3ex\hbox{$<$}\mkern-14mu
             \lower0.6ex\hbox{$\sim$}}}

\newcommand{\bL}{\bm L}
\newcommand{\bS}{\bm S}
\newcommand{\bJ}{\bm J}
\newcommand{\bH}{\bm H}
\newcommand{\bl}{\bm {\hat l}}
\newcommand{\bs}{\bm {\hat s}}

\newcommand{\der}{\text{d}}
\newcommand{\tildea}{{\tilde a}}

\def\red#1 {\textcolor{red}{#1}\ }   
\def\blue#1 {\textcolor{blue}{#1}\ }   

\begin{document}
\title[Spin-Orbit Misalignment]
{How do External Companions Affect Spin-Orbit Misalignment of Hot Jupiters?}
\author[D.~Lai et al]{Dong Lai\thanks{Email: dong@astro.cornell.edu},
Kassandra R. Anderson and Bonan Pu\\
Cornell Center for Astrophysics and Planetary Science, Department of Astronomy,
Cornell University, Ithaca, NY 14853, USA\\}

\pagerange{\pageref{firstpage}--\pageref{lastpage}} \pubyear{2013}

\label{firstpage}
\maketitle

\begin{abstract}
Consider a planet with its orbital angular momentum axis aligned with
the spin axis of its host star. To what extent does an inclined distant
companion (giant planet or binary star) affect this alignment? We
provide an analytic, quantitative answer and apply it to hot Jupiter
systems, for which misalignments between the orbital axis and the
stellar spin axis have been detected. We also show how similar
consideration can be applied to multi-planet systems with distant
companions (such as Kepler-56). The result of this paper provides a
simple method to assess the dynamical role played by external
companions on spin-orbit misalignments in exoplanetary systems.
\end{abstract}


\section{Introduction}

Many hot Jupiters (HJs, giant planets with orbital periods about 
5 days or less) are known to have a misaligned orbital angular momentum
axis with respective to the spin axis of their host stars (e.g. Winn
\& Fabrycky 2015).  The origin of this spin-orbit misalignment and the
formation mechanisms of HJs remain under debate. The possibilities
include high-eccentricity migration induced by additional giant
planets or stellar companions (e.g., Fabricky \& Tremaine 2007;
Wu et al.~2007; Nagasawa et al.~2008; Wu \& Lithwick 2011; Naoz et al.~2012; Beauge \&
Nesvorny 2012; Storch et al.~2014; Petrovich 2015; Anderson et al.~2016; Munoz et
al.~2016) and migration in protoplanetary disks that are misaligned
with the host stars (e.g. Bate et al.~2010; Lai et al.~2011; Batygin
2012; Batygin \& Adams 2013; Lai 2014; Spalding \& Batygin 2014).

A large fraction ($\sim 70\%$) of HJ systems have been found to have
external companions, either gas giants at large semi-major axes
(1-20~au) (Knutson et al.~2014; Bryan et al.~2016) or distant
(50-2000~au) stellar companions (Ngo et al.~2015; Wang et al.~2015).
How these companions might have facilitated the formation of
misaligned HJs is an important question, but often may not have an
unambiguous answer (e.g. Ngo et al.~2016; Atonini et al.~2016). 
A simpler question to ask is how an observed
companion may affect the current state of spin-orbit misalignment. The
answer to this question, when combined with observational data, may
help constrain the property of the companion.  The purpose of this
paper is to provide such an answer in quantitative way (Section 2), to
allow quick assessment/constraint of the dynamical role of external
companions on spin-orbit misalignments, without formal celestial
mechanics calculations (e.g. Boue \& Fabrycky 2014)
or N-body integrations with large parameter
space. In Section 3, we illustrate our result by applying it to several HJ
systems with external companions. A recent work (Becker et al.~2017)
did not include the gravitational coupling between the HJ and the oblate
host star; this coupling can in fact play an important role in determining 
how spin-orbit misalignments are affected by the companion.

Spin-orbit misalignment has been detected in at least one multi-planet
system (Kepler-56; Huber et al.~2013). Our analysis can be generalized
to such systems. This is discussed in Section 4 using Kepler-56 as
an example.

\section{Equations and Analytical Results}

Consider a planet of mass $m_1$ on a circular orbit (with semi-major
axis $a_1$) around a central star (mass $M_\star$ and spin angular
momentum vector $\bS_\star$). The orbital angular momentum vector of
the planet is $\bL_1$. An external perturber (mass $m_p$) moves on an
inclined orbit, with semi-major axis $a_p$, eccentricity $e_p$ and
inclination $\theta_p$ (the angle between $\bL_1$ and $\bL_p$, the
orbital angular momentum of the perturber).  How is the spin-orbit
misalignment angle $\theta_{\star 1}$ (the angle between $\bS_\star$
and $\bL_1$ influenced by the perturber?

We denote the relevant (spin and orbital) angular momentum vectors by
$\bS_\star =S_\star\bs_\star$, $\bL_1=L_1\bl_1$ and $\bL_p=L_p\bl_p$, where 
$\bs_\star$, $\bl_1$ and $\bl_p$ are unit vectors. 
The evolution equations for $\bs_\star$, $\bl_1$ and $\bl_p$ are
\ba
&&{\der\bs_\star\over \der t}=\omega_{\star 1}(\bl_1\cdot\bs_\star)(\bs_\star\times\bl_1)
+\omega_{\star p}(\bs_\star\cdot\bl_p)(\bs_\star\times\bl_p),\label{eq:ds}\\
&&{\der\bl_1\over \der t}=\omega_{1\star}(\bl_1\cdot\bs_\star)(\bl_1\times\bs_\star)
+\omega_{1p}(\bl_1\cdot\bl_p)(\bl_1\times\bl_p),\label{eq:dl1}\\
&&{\der\bl_p\over \der t}=\omega_{p\star}(\bl_p\cdot\bs_\star)(\bl_p\times\bs_\star)
+\omega_{p1}(\bl_p\cdot\bl_1)(\bl_p\times\bl_1).\label{eq:dlp}
\ea
Each term in the above equations has a clear physical meaning.
The characteristic precession rate of $\bs_\star$ around
$\bl_1$ (driven by $m_1$) is given by 
\be
\omega_{\star 1}={3k_{q\star}\over 2k_\star}\left( {m_1\over M_\star}\right)
\left({R_\star\over a_1}\right)^3\Omega_\star,\label{eq:os1}
\ee
where $\Omega_\star=2\pi/P_\star$ is the angular frequency of the host star,
and $k_\star$,$k_{q\star}$ are defined through the star's moment of inertia and
quadrupole moment: $I_3=k_\star M_\star R_\star^2$ and $I_3-I_1                 
=k_{q\star}{\hat\Omega}_\star^2 M_\star R_\star^2$, with $\hat\Omega_\star=     
\Omega_\star (GM_\star/R_\star^3)^{-1/2}$.
Typical values (for solar type stars) are
$k_\star\simeq 0.06$ and $k_{q\star}\simeq 0.01$
(Mecheri et al.~2004; Lai 2016).
Similarly, the characteristic precession rate of $\bs_\star$ around
$\bl_p$ (driven by $m_p$) is given by 
\be
\omega_{\star p}={3k_{q\star}\over 2k_\star}\left( {m_p\over M_\star}\right)
\left({R_\star\over \tildea_p}\right)^3\Omega_\star,\label{eq:osp}
\ee
where we have defined the ``effective'' semi-major axis 
$\tildea_p\equiv a_p\sqrt{1-e_p^2}$.
The characteristic precession rate of $\bl_1$ around $\bs_\star$ 
(driven by the stellar quadrupole) is given by 
\be
\omega_{1 \star}=\omega_{\star 1} \left({S_\star\over L_1}\right).
\label{eq:o1s}
\ee
The characteristic precession rate of $\bl_1$ around $\bl_p$ (driven by $m_p$)
is given by 
\be
\omega_{1p}={3m_p\over 4M_\star}\left({a_1\over \tildea_p}\right)^3 n_1,
\label{eq:o1p}
\ee
where $n_1$ is the mean motion of $m_1$. The two precession frequencies characterizing
the evolution of $\bl_p$ are
\be
\omega_{p\star}=\omega_{\star p} \left({S_\star\over L_p}\right),\quad
\omega_{p1}=\omega_{1p} \left({L_1\over L_p}\right).
\label{eq:op}
\ee
Obviously, when $L_p\gg L_1$ and $L_p\gg S_\star$, we can consider $\bl_p$ fixed in time.

The above assumes a circular orbit for $m_1$. An inclined perturber may
also excite eccentricity. But this is suppressed by various
short-range forces for close-in planets (e.g. Liu et al.~2015): For
example, the pericenter precession rate due to General Relativity,
$(3GM_\star/c^2a_1)n_1$, is much larger than the Lidov-Kozai rate,
$(m_p/M_\star)(a_1/\tildea_p)^3 n_1$, for systems of interest in this
paper.

\subsection{Results}

\begin{figure}
\centering
\includegraphics[scale=0.57]{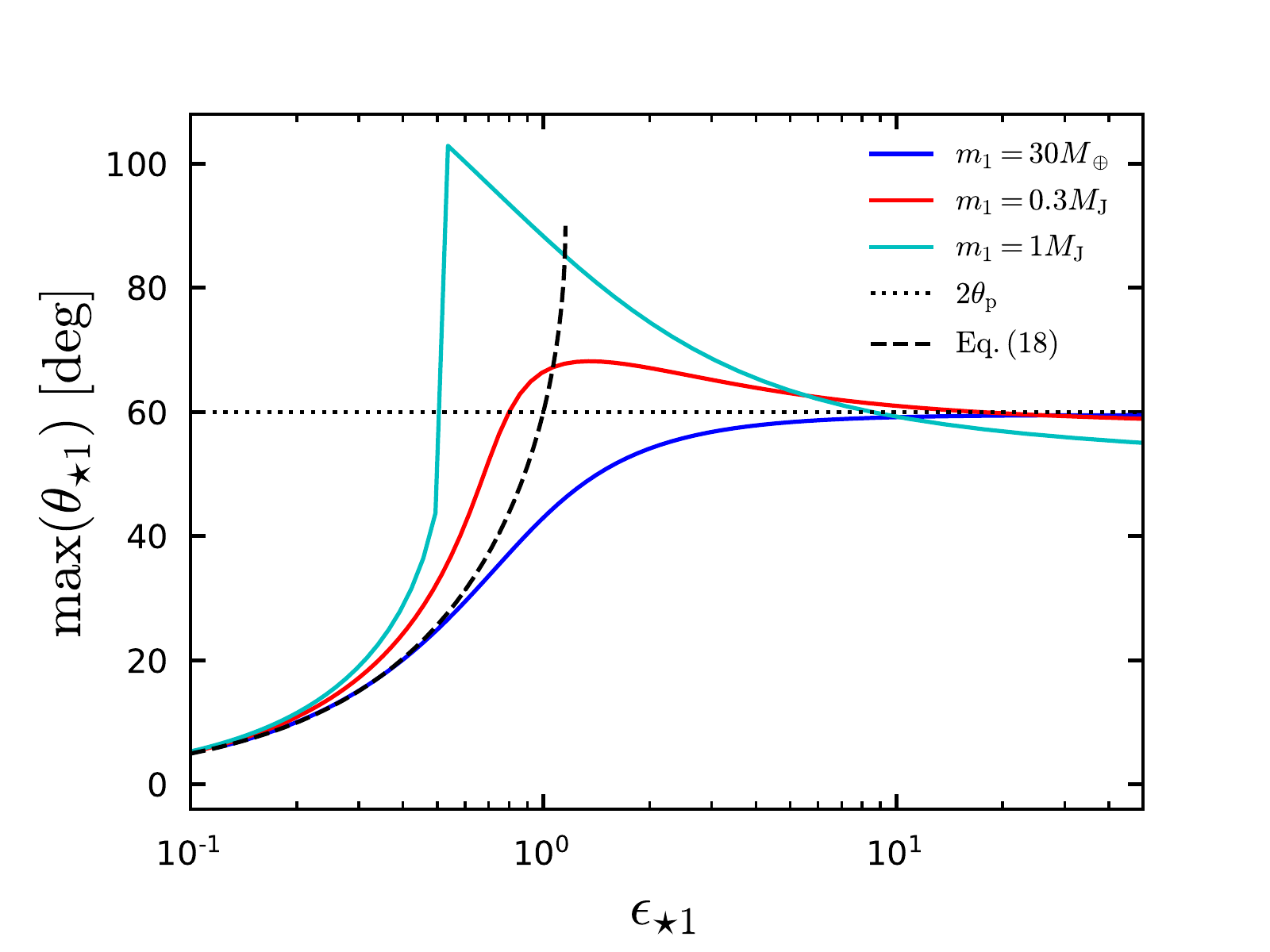}
\caption{Maximum misalignment angle between $\bl_1$
and $\bs_\star$ 
in the presence of an external inclined perturber ($m_p$). Initially 
$\bl_1$ and $\bs_\star$ are aligned, and both inclined relative to $\bl_p$ 
at $\theta_p=30^\circ$. We fix the HJ and stellar parameters 
to those of WASP-41 (see Section 3), but allow $m_1$ to
take various values.  Different solid lines correspond to 
different values of $m_1$ ($=30 M_{\oplus}$, $0.3 M_{\rm J}$, $1
M_{\rm J}$, corresponding to $L_1/S_\star= 0.9, 2.9, 9.7$).
The dimensionless coupling parameter
$\epsilon_{\star 1}$ (Eq.~\ref{eq:epss1})
is varied by varying the ``strength'' of the perturber,
$m_p/\tildea_p^3$. Analytical results in the strong (planet-star) coupling and weak
coupling limits are also shown. A resonance feature is present around $\epsilon_{\star 1}
\sim 1$ when $L_1/S_\star\gtrsim 1$. Note that in general
$(\theta_{\star 1})_{\rm max}$ depends on various physical parameters through 
$\epsilon_{\star 1}$, $L_1/S_\star$ and $\theta_p$ (and $L_p/L_1$ when it is not 
much larger than unity).}
\label{fig1}
\end{figure}

Given initial conditions, equations (\ref{eq:ds})-(\ref{eq:dlp}) can be easily integrated to
determine the time evolution of $\theta_{\star 1}$ and its maximum value.
Figure \ref{fig1} shows some examples of $(\theta_{\star 1})_{\rm max}$
for initially spin-orbit aligned systems ($\theta_{\star 1}=0$, but $\theta_p\neq 0$)
with various parameters.

The problem we consider here is analogous to the problem of the
evolution of the mutual inclination of two planets under the influence
of an inclined, distant perturber (Lai \& Pu 2017). A key dimensionless
parameter is
\be 
\epsilon_{\star 1}\equiv
{\omega_{1p}-\omega_{\star p}\over\omega_{\star 1}+\omega_{1\star}},
\label{eq:epss1}
\ee
which measures the strength of planet-star coupling relative to the
``disruptive'' forcing of the perturber (planet-star coupling
increases with decreasing $\epsilon_{\star 1}$).
While there are many physical parameters in this problem 
($m_1,~m_p,~a_1,~a_p$, stellar spin, etc),
the shape of the $(\theta_{\star 1})_{\max}-\epsilon_{\star 1}$ curve shown in 
Fig.~\ref{fig1} (for initial aligned systems) depends only on $L_1/S_\star$ and $\theta_p$
(and somewhat on $L_p/L_1$ when $L_p$ is only moderately larger than $L_1$).

There are three regimes for the behaviors of $\theta_{\star 1}$ evolution:

(i) {\it Weak (planet-star) coupling regime} ($\epsilon_{\star 1}\gg 1$). 
In this regime, both $\bl_1$ and $\bl_p$ precess around $\bJ=\bL_1+\bL_p=J{\hat\bJ}$
at a rapid rate:
\be
{\der\bl_{1,p}\over \der t}\simeq \left({J\omega_{1p}\over L_p}\cos\theta_p\right)\,
\bl_{1,p}\times {\hat\bJ},
\ee
with a constant $\theta_{1J}$ (the angle between $\bl_1$ and $\bJ$) given by
$\sin\theta_{1J}=(L_p/J)\sin\theta_p$,
where $J=|\bJ|=(L_1^2+L_p^2+2L_1L_p\cos\theta_p)^{1/2}$. On the other hand, $\bs_\star$ 
also precesses around $\bJ$, but at a much slower rate:
\be
{\der\bs_\star\over \der t}\simeq \left(\omega_{\star 1}\cos^2\theta_{1J}+\omega_{\star p}
\cos^2\theta_{pJ}\right)(\cos\theta_{\star J})\, \bs_\star\times {\hat\bJ},
\ee
with a (approximately) constant $\theta_{\star J}$ (the angle between 
$\bs_\star$ and $\bJ$). (Of course, $\bs_\star$ and $\bJ$ mutually precess, conserving the 
total angular momentum.) Thus $\theta_{\star 1}$ varies between a minimum and maximum 
given by 
\ba
&&\left(\theta_{\star 1}\right)_{\rm min}\simeq |\theta_{\star J}-\theta_{1J}|,\label{eq:tmin}\\
&&\left(\theta_{\star 1}\right)_{\rm max}\simeq \theta_{\star J}+\theta_{1J}.\label{eq:tmax}
\ea

For an initially aligned ($\theta_{\star 1}=0$) system, we have
\be
\left(\theta_{\star 1}\right)_{\rm max}
\simeq 2\theta_{1J}=2\sin^{-1}\!\left({L_p\over J}\sin\theta_p\right).
\ee
For $L_p\gg L_1$, this reduces to $(\theta_{\star 1})_{\rm max}=2\theta_p$.

(ii) {\it Strong (planet-star) coupling regime} ($\epsilon_{\star 1}\ll 1$).
In this regime, $\bs_\star$ and $\bl_1$ precess around each other at a rapid rate,
preserving the misalignment angle as an adiabatic invariant
\footnote{This result was first derived by Goldreich (1965) for $S_\star\gg L_1$
(in the context of planetary satellites, with the Sun acting as a perturber).
The generalization to comparable $S_\star$ and $L_1$ is straightforward:
Both $\bS_\star$ and $\bL_1$ precess rapidly around $\bH$, while $\bH$ precesses
around $\bL_p$ slowly; thus $\cos\theta_{\star H}$ is adiabatically invariant
because it is the ``action'' of a action-angle pair.}
\be
\theta_{\star 1}\simeq {\rm constant}.
\label{eq:ts1}
\ee
The angle between $\bL_1$ and $\bH=\bS_\star+\bL_1$ is also constant, and is given by 
$\sin\theta_{1H}=(S/H)\sin\theta_{\star 1}$.  On the other hand, $\bL_p$ and $\bH$
mutually precess around each other at a slow rate, with a constant $\theta_{pH}$ (the angle
between $\bL_p$ and $\bH$). Thus, in the strong coupling regime, $\theta_{1p}=\theta_p$
is not constant, but varies between 
a minimum and maximum given by 
\ba
&&\left(\theta_{1p}\right)_{\rm min}\simeq |\theta_{pH}-\theta_{1H}|,\label{eq:pmin}\\
&&\left(\theta_{1p}\right)_{\rm max}\simeq \theta_{pH}+\theta_{1H}.\label{eq:pmax}
\ea

For initially aligned systems ($\theta_{\star 1}=0$), $\theta_p$ is approximately 
constant (since $\theta_{1H}=0$). 
We can obtain the leading-order correction to the adiabatic result
($\theta_{\star 1}=0$; see Eq.~\ref{eq:ts1}):
the maximum and the RMS values of $|\sin\theta_{\star 1}|$ are given by 
(see Lai \& Pu 2017)
\ba
&&\left|\sin\theta_{\star 1}\right|_{\rm max}\simeq \epsilon_{\star 1}
\left|\sin 2\theta_p\right|, \label{eq:th12max}\\
&&\left\langle\sin^2\!\theta_{\star 1}\right\rangle^{\!1/2}\simeq
{1\over\sqrt{2}}\epsilon_{\star 1}\left|\sin 2\theta_p\right|.
\label{eq:th12max2}
\ea
Note that these equations are valid for arbitrary values of $\theta_p$.

(iii) {\it Resonance.}
For $\epsilon_{\star 1}\sim 1$, resonant excitation of
$\theta_{\star 1}$ becomes possible. More precisely, for small $\theta_{\star 1}$
and $\theta_p$, resonance occurs when 
\be
\omega_{\star 1}+\omega_{\star p}\simeq \omega_{1\star}+\omega_{1p},
\ee
which is equivalent to
\be
\epsilon_{\star 1}\simeq {1-S_\star/L_1\over 1+S_\star/L_1}.
\ee
Obviously, since $\omega_{1p}$ is always larger than $\omega_{\star p}$,
resonance is possible only when $S_\star\lesssim L_1$. Finite $\theta_p$ tends
to broaden or smooth out the resonance, although a significant ``discontinuity''
in the $(\theta_{\star 1})_{\rm max}-\epsilon_{\star 1}$ curve is still visible for
$L_1/S_\star\gtrsim$~a few [Fig.~\ref{fig1}. See Appendix A of Lai \& Pu (2017) for 
more details on the theory of resonance].

The above results/discussions are valid for arbitary $\theta_p$. For $\theta_p\ll 1$,
an analytical expression of $(\theta_{\star 1})_{\rm max}$ covering 
all three regimes can be obtained (Pu \& Lai 2017).

\section{Hot Jupiter Systems with External Companion}

For a HJ around a solar-type star ($M_\star\sim M_\odot,~R_\star\sim R_\odot$)
with a giant planet ($m_p\sim M_J$) perturber at $\tildea_p\sim 1$~au, we have
\ba
&& \omega_{\star 1}\simeq 4.8\times 10^{-6}
\left({6k_{q\star}\over k_\star}\right){m_1\over M_J}\left({a_1\over 0.04\,{\rm au}}\right)^{\!-3}
\left(\frac{M_\star}{M_\odot} \right)^{-1} \nonumber\\
&& \qquad \times \left(\frac{R_\star}{R_\odot}\right)^{\!3}\!
\left({P_\star\over 30\,{\rm d}}\right)^{\!-1}{2\pi\over {\rm yrs}},\label{eq:eq20}\\
&& \omega_{1p}\simeq 6\times 10^{-6}\,
{m_p\over M_J}\left(\frac{M_\star}{M_\odot} \right)^{\!\!-1/2}\!\! 
\left({a_1\over 0.04\,{\rm au}}\right)^{\!3/2}\nonumber\\
&&\qquad \times \left({\tildea_p\over 1\,{\rm au}}\right)^{\!-3}{2\pi\over {\rm yrs}},
\label{eq:eq21}\ea
where $6k_{q\star}/k_\star\simeq 1$. Thus
\ba
&&\epsilon_{\star 1}={\omega_{1p}\over\omega_{\star 1}}
\left({1-\omega_{\star p}/\omega_{1p}\over 1+S_\star/L_1}\right)\nonumber\\
&&\quad ~\simeq 1.25\,
\left({6k_{q\star}\over k_\star}\right)^{\!\!-1}\,{m_p\over m_1}\,
\left({a_1\over 0.04\,{\rm au}}\right)^{\!9/2}
\left({\tildea_p\over 1\,{\rm au}}\right)^{\!-3}
\left({P_\star\over 30\,{\rm d}}\right)\nonumber\\
&&\qquad ~~\times \left(\frac{M_\star}{M_\odot}\right)^{\!1/2} 
\left(\frac{R_\star}{R_\odot} \right)^{\!\!-3} \left( {1\over 1+S_\star/L_1} \right),
\label{eq:eq22}\ea
where in the second equality we have used 
$\omega_{\star p}/\omega_{1p}\ll 1$, and the angular momentum ratio is
\ba
&& {S_\star\over L_1}=0.079\left({k_\star\over 0.06}\right)\!\left(\!
{m_1\over M_J}\!\right)^{\!\!-1}\!\left({a_1\over 0.04\,{\rm au}}\right)^{\!-1/2}
\!\left({P_\star\over 30\,{\rm d}}\right)^{\!-1} \nonumber\\
&& \qquad \times \left( \frac{M_\star}{M_\odot} \right)^{\!\!1/2} 
\left(\frac{R_\star}{R_\odot} \right)^2.
\label{eq:eq23}\ea

\begin{figure}
\centering
\includegraphics[scale=0.5]{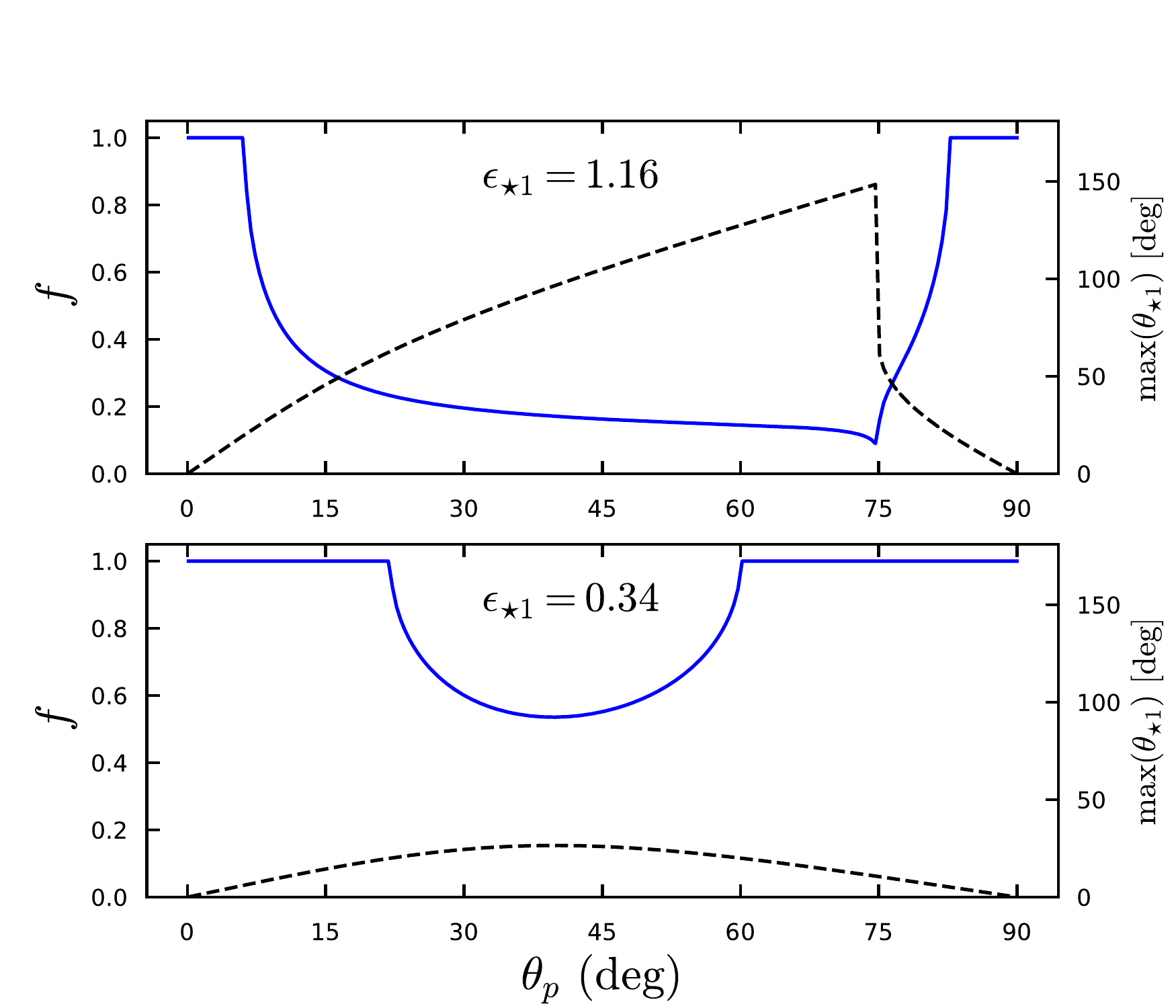}
\caption{A typical HJ system with $m_1 = 1 M_{\rm J}$, $a_1 = 0.04$~au, 
orbiting a star with $M_\star = 1 M_\odot$, $R_\star = 1
  R_\odot$, $P_\star = 30$ days.  An external companion has $m_p = 1
  M_{\rm J}$, $a_p = 1 $~au (top panel), and $a_p = 1.5$~au (bottom
  panel), leading to differing values of $\epsilon_{\star 1}$, as
  labeled.  Considering a low-obliquity HJ with projected obliquity
  $\lambda$ constrained to be in the range $0 - 20^{\circ}$, we calculate the
  fraction of time $f$ that $\theta_{\star 1}$ is within that range as
  a function of $\theta_p$.  The dashed lines depict the maximum value
  of $\theta_{\star 1}$ achieved over the integration timespan (1~Gyr).
  Note that in the upper panel, the maximum $\theta_{\star 1}$ for large $\theta_p$ 
is not equal to the theoretical $(\theta_{\star 1})_{\rm max}$ discussed in Section 2 
(see Fig.~\ref{fig1}); this is because for large $\theta_p$, the precession 
time of $\bl_1$ around $\bl_p$ can be longer than the integration time, so the system
does not have time to reach the theoretical maximum $\theta_{\star 1}$.}
\label{fig2}
\end{figure}

\subsection{Constrain External Perturbers of HJs Using Stellar Obliquities}

Radial velocity detections of external companions of HJs give $m_p\sin i$ and
$a_p,~e_p$, but do not constrain $\theta_p$. Rossiter-McLaughlin
measurements yield the projected stellar obliquity $\lambda$, which is related to
the 3D stellar obliquity $\theta_{\star 1}$ by
\be
\sin^2\theta_{\star 1}={\sin^2\lambda \over 1-\cos^2\lambda\cos^2\phi},
\label{eq:true}\ee
where $\phi$ is defined by $\bs_\star=
\sin\theta_{\star 1}({\hat {\bf x}}\cos\phi + {\hat{\bf y}}
\sin\phi)+\bl_1\cos\theta_{\star 1}$, with ${\hat{\bf x}}$ along the line of sight.
Several systems containing a HJ and external companion were recently
considered by \cite{becker2017} in an attempt to constrain $\theta_p$, but 
they neglected the stellar spin-orbit coupling in their analysis.

In the following, we do not attempt to carry out full-blown
statistical analysis.  Instead, for each system, we evaluate the
coupling parameter $\epsilon_{\star 1}$, which immediately informs us
whether the observed stellar obliquity is ``permanent'' or can vary due
to an inclined perturber. In addition, assuming an initial $\theta_{\star 1}=0$,
we calculate the fraction of
time $f$ that the system spends with $\theta_{\star 1}$ less than certain
value (as constrained by observations) for a range of $\theta_p$'s.
This would then provide an approximate constraint on $\theta_p$.

Figure \ref{fig2} illustrates our procedure. It highlights the
importance of including spin-orbit coupling in the analysis and the
role of the coupling parameter $\epsilon_{\star 1}$. In particular,
for $\epsilon_{\star 1}\lesssim 0.4$, even a highly inclined perturber
would satisfy the observational constraint of small $\lambda$.

\subsection{Specific Systems}

\subsubsection{HAT-P-13}

\begin{figure}
\centering
\includegraphics[scale=0.55]{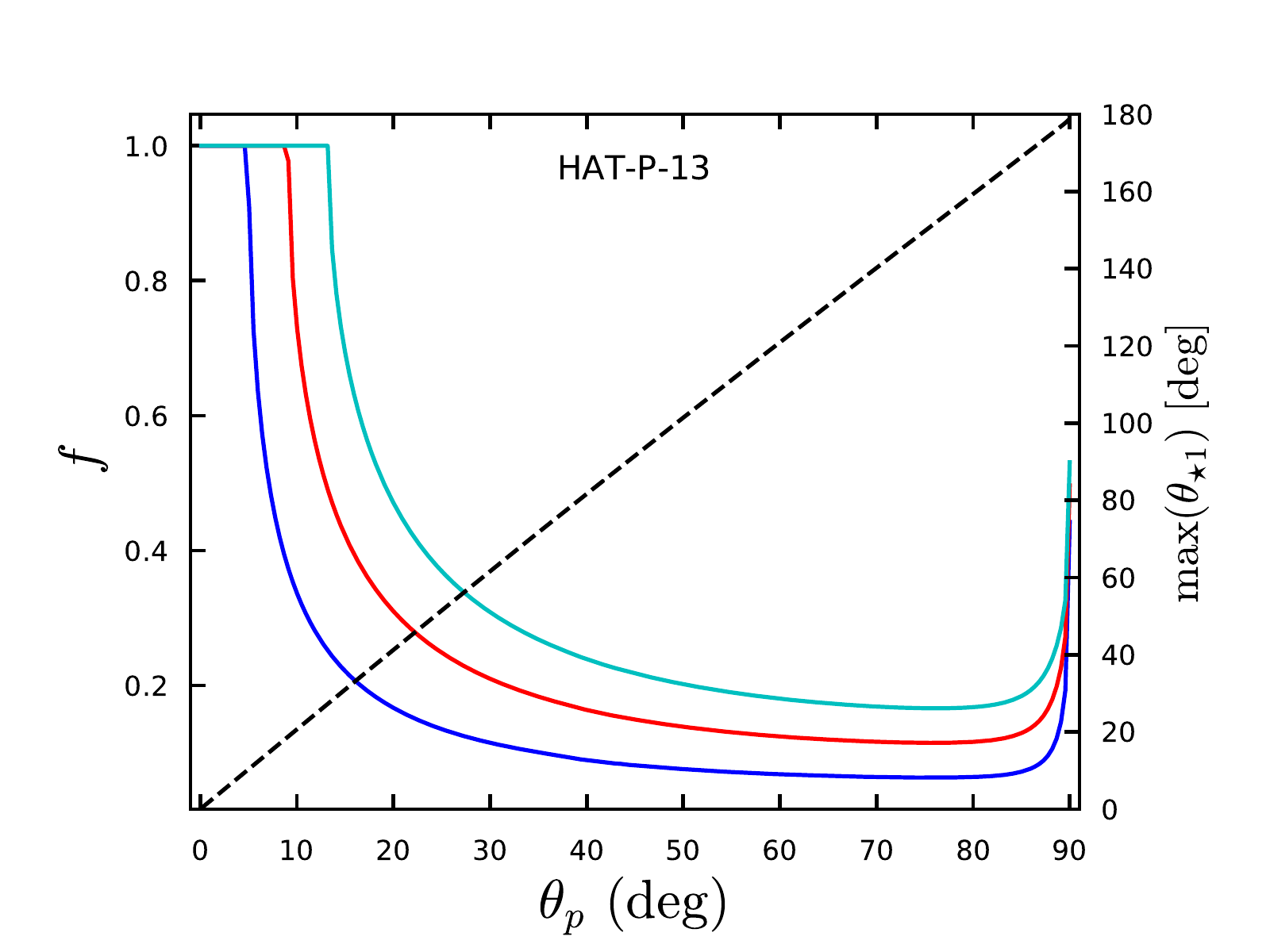}
\caption{Similar to Fig.~\ref{fig2} but showing results for
  HAT-P-13. Fraction of time $f$ with $\theta_{\star 1}$ spent in the
  observed range of obliquity, as a function of $\theta_p$. We use the projected
  obliquity $\lambda$ as an indicator of the range of true obliquity 
  (which is generally larger than $\lambda$; see Eq.~\ref{eq:true}).
  Blue (lower solid curve): $\theta_{\star 1}$ within $1 \sigma$ of the observed $\lambda$;
  red (middle solid curve): $2 \sigma$; cyan (upper curve): $3 \sigma$.
  Dashed line: maximum value of $\theta_{\star 1}$ achieved over the
  integration timespan of 1 Gyr.}
\label{fig3}
\end{figure}

This system ($M_\star = 1.22^{+0.05}_{- 0.10} M_\odot$, $R_\star = 1.56\pm 0.08 R_\odot$)
contains a HJ with $m_1 = 0.851\pm 0.038 M_{\rm J}$, $a_1 = 0.0427^{+0.0006}_{-0.0012}\ {\rm au}$, 
and an external perturber with $m_p \sin i= 14.28\pm 0.28M_{\rm J}$, 
$a_p = 1.226 \ {\rm au} $, $e_p = 0.6616\pm 0.0054$
\citep{bakos2009, winn2010}.  The star has a measured $v \sin i_\star = 
1.66\pm 0.37\ {\rm km/s}$, yielding a stellar rotation period $P_\star \simeq 47.5\sin i_\star
\ {\rm days} $.  The projected spin-orbit misalignment angle is 
$\lambda=1.9^\circ\pm 8.5^\circ$.
For concreteness, we adopt the measured mean values for various 
quantities and assume $\sin i=\sin i_\star=1$
(these loose assumptions will also be used for the other systems considered below).
We then find $\epsilon_{\star 1} \simeq 14$. (Note that a larger $m_p$ would increase 
$\epsilon_{\star 1}$, while a smaller $P_\star$ would decrease 
$\epsilon_{\star 1}$.) The system is therefore safely in the 
weak-coupling regime, and may have its spin-orbit alignment disrupted if the
perturber is sufficiently inclined (see Fig.~\ref{fig3}).

\subsubsection{WASP-41}

\begin{figure}
\centering
\includegraphics[scale=0.55]{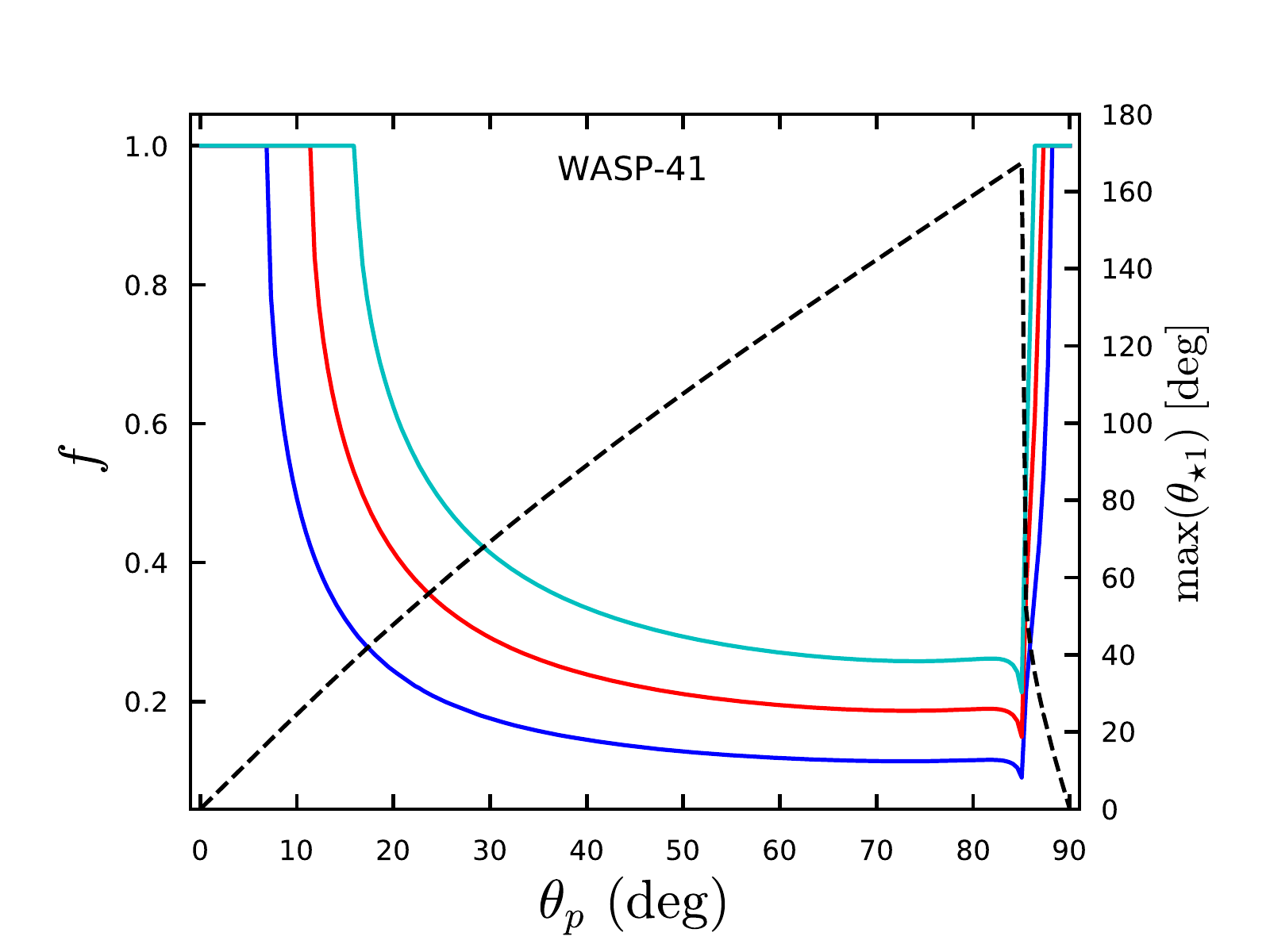}
\caption{Same as Fig.~\ref{fig3} but showing results for WASP-41.}
\label{fig4}
\end{figure}

This system ($M_\star = 0.93\pm 0.07 M_\odot$, $R_\star = 0.87\pm 0.03 R_\odot$) 
contains a HJ with $m_1 = 0.94\pm 0.05 M_{\rm J}$, $a_1 = 0.040\pm 0.001
\ {\rm au}$, and an external perturber with $m_p\sin i = 3.18\pm 0.2 M_{\rm J}$,
$a_p = 1.07\pm 0.03\ {\rm au}$, $e_p = 0.294\pm 0.024$ \citep{neveu2016}.  
RM measurement gives $\lambda=6^\circ\pm 11^\circ$.
The star has $v \sin i_\star = 2.64\pm 0.25 \ {\rm km/s}$, yielding $P_\star \simeq 
16.7\sin i_\star\ {\rm days}$. Adopting the measured mean values for various
quantities and assuming $\sin i=\sin i_\star=1$,
we find $\epsilon_{\star,1} \simeq 2.9$. This system is in the weak coupling regime.
The $\theta_p$ constraint is similar to that for HAT-P-13 (see Fig.~\ref{fig4}).

\subsubsection{WASP-47}

\begin{figure}
\centering
\includegraphics[scale=0.55]{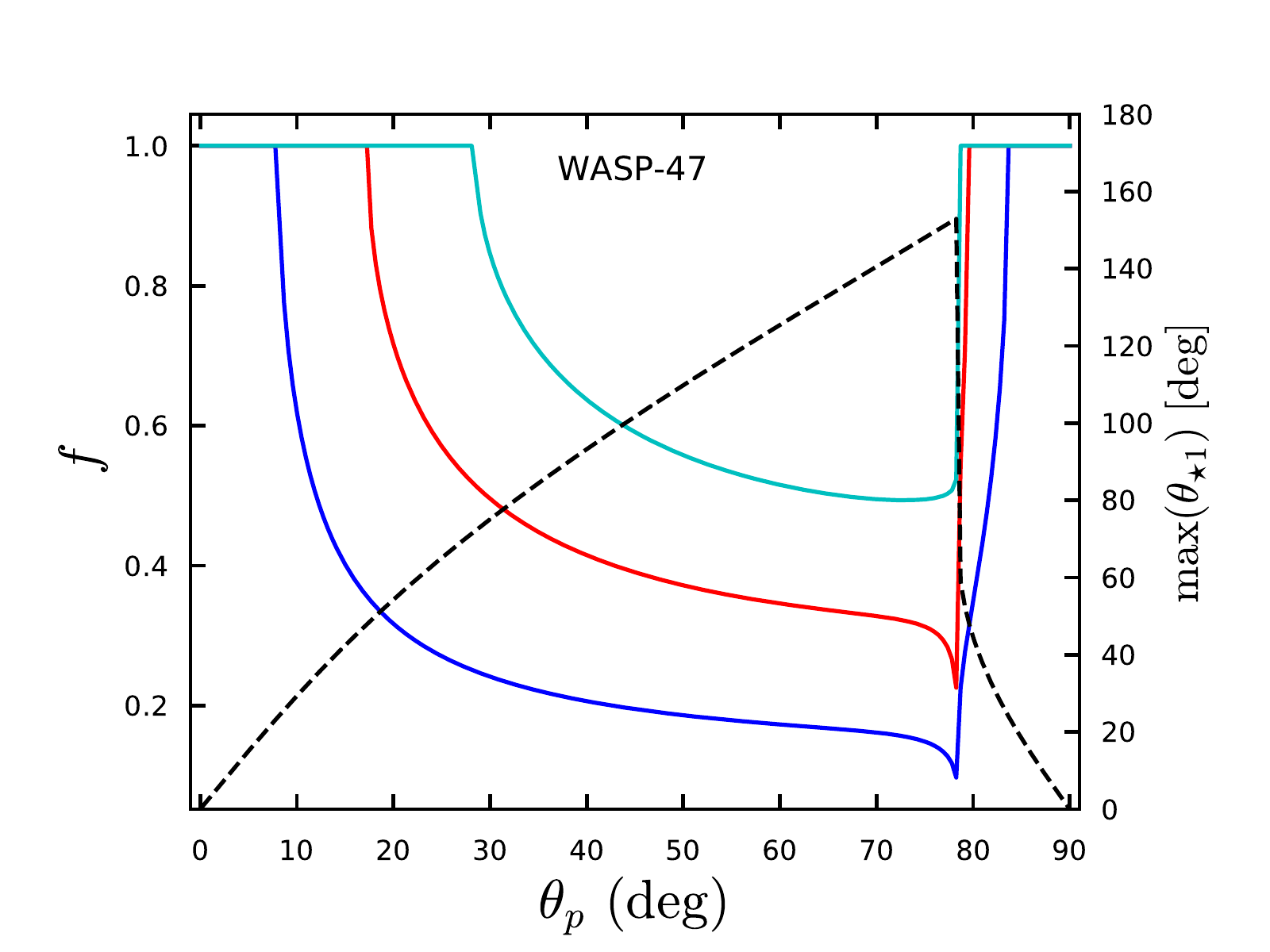}
\caption{Same as Fig.~\ref{fig3} but showing results for WASP-47.}
\label{fig5}
\end{figure}

This system ($M_\star = 1.026\pm 0.076 M_\odot$, $R_\star = 1.15 \pm
0.04R_\odot$) contains a HJ with $m_1 = 1.13\pm 0.06 M_{\rm J}$, $a_1
= 0.051\pm 0.001 \ {\rm au}$, and an external perturber with $m_p\sin
i = 1.24\pm 0.22M_{\rm J}$, $a_p = 1.36\pm 0.04 \ {\rm au}$, $e_p =
0.13\pm 0.1$ \citep{neveu2016}.  The HJ also has two close
low-mass neighbors \citep{becker2015}, whose dynamical effects are
negligible for this analysis.  RM measurement gives $\lambda=0\pm
24^\circ$ (Sanchis-Ojeda et al.~2015) The star has $v \sin i_\star =
1.3\pm 1 \ {\rm km/s}$ \citep{sanchis2015} or $3.0\pm 0.6 \ {\rm
  km/s}$ (Hellier et al.~2012).  Choosing $v \sin i_\star = 1.3 {\rm
  km/s}$ yields $P_\star \simeq 44.8\sin i_\star \ {\rm days}$.  As a result,
$\epsilon_{\star,1} \simeq 1.6$, putting the system in the weak coupling regime.
Note that $\epsilon_{\star 1}$ can be easily smaller by 
a factor of two or more, given the uncertainty in $P_\star$. Overall, because of the
large error in the $\lambda$ measurement, $\theta_p$ is not well constrained
(see Fig.~\ref{fig5}).

\subsubsection{WASP-22}

This system ($M_\star = 1.109 \pm 0.026M_\odot$, $R_\star =
1.219^{+0.052}_{-0.033} R_\odot$) contains a HJ with $m_1 = 0.588 \pm
0.017 M_{\rm J}$, $a_1 = 0.04698 \pm 0.00037\ {\rm au}$, and observed
linear RV trend of $\sim 40$ m/s/yr, indicating the presence of an
external perturber \citep{maxted2010, anderson2011}.  The projected
mass and semi-major axis of the perturber are constrained in the
combination $m_p \sin i/a_p^2$ \citep[see
Fig.~10 of][]{knutson2014}.  RM measurement gives
$\lambda=22^\circ \pm 16^\circ$.  For a planetary or brown dwarf
($m_p\lesssim 30 M_J$) companion, this system is always in the
strong-coupling regime with $\epsilon_{\star 1} \ll 1$, 
so that the perturber is unable to change the spin-orbit angle from
the initial value for any value of $\theta_p$.
Whatever the observed value 
$\theta_{\star 1}$ must be ``permanent'', and no meaningful constraint on $\theta_p$ 
can be obtained (see the lower panel of Fig.~\ref{fig2}).

\section{Spin-Orbit Misalignment in Multi-planet Systems: Kepler-56}

Our analysis can be generalized to multi-planet systems with spin-orbit 
misalignment measurements. Here we use Kepler-56 as an example to illustrate
our method, as this is the only such system with
significant stellar obliquity (Huber et al.~2013).

Kepler-56 (with a red giant host star $M_\star=1.32M_\odot$,
$R_\star=4.23R_\odot$, $P_\star=74$~days) has two transiting planets
($m_{1,2}=0.0925,\,0.60M_J$) at $a_{1,2}=0.103,\,0.165$~au (period
$10.5,\,21$~days). The orbits of the two planets are coplanar within
$\sim R_\star/a_2=6.8^\circ$, and are inclined with respect to the
stellar equator by more than $37^\circ$ (Huber et al.~2013).  An
external perturber has been found through RV observations, with
$a_p=2.15$~au, $e_p=0.2$, and $m_p\sin i\simeq 5.6M_J$ (Otor et al.~2016).

The mutual coupling between the two inner planets against the
forcing from the perturber can be measured by the parameter (Lai \& Pu 2017)
\be
\epsilon_{12}\equiv {\omega_{2p}-\omega_{1p}\over\omega_{12}+\omega_{21}},
\ee
where the $\omega$'s have the similar meanings as in Section 2.
We find $\epsilon_{12}\simeq  2.3\times 10^{-3}/\sin i$, implying that
the inner two planets are strongly coupled and their coplanarity is
not affected by any external perturbers that satisfy the current RV constraint
(regardless of $\theta_p$). It is also easy to check that the two planets are
strongly coupled with respect to the perturbation from the stellar quadrupole.

Thus, the two inner planets tend to precess around $\bl_p$ as a ``rigid'' body, with
characteristic frequency
\be
\omega_{12,p}={L_1\omega_{1p}+L_2\omega_{2p}\over L_1+L_2}\simeq {2\pi\over 4.3\times 10^4\
{\rm yrs}}\left({m_p\over 5.6M_J}\right).
\ee
The two inner planets also drive the star into precession, with characteristic frequency
\be
\omega_{\star,12}=\omega_{\star 1}+\omega_{\star 2}\simeq {2\pi\over 6.7\times 10^5\,
{\rm yrs}}\left({6k_{q\star}\over k_\star}\right).
\ee
The precession of $\bl_1\simeq \bl_2$ around $\bs_\star$ is
\be
\omega_{12,\star}=\left({S_\star \over L_1+L_2}\right)
\omega_{\star,12}\simeq 0.49\left({k_\star\over 0.06}
\right)\omega_{\star,12}.
\ee
Thus the coupling parameter between $\bs_\star$ and $\bl_{1,2}$ relative to the forcing from
$m_p$ is
\be
\epsilon_{\star,12}={\omega_{12,p}-\omega_{\star,p}\over \omega_{\star,12}
+\omega_{12,\star}}\simeq 15.6\left({m_p\over 5.6M_J}\right){k_\star/(6k_{q\star})\over 
1+8k_\star}.
\label{eq:epss12}\ee
Even with the uncertainties of various parameters, the inner two planets are weakly 
coupled to the stellar spin. The stellar obliquity (relative to $\bl_{1,2}$) therefore 
varies between $|\theta_{\star p}-\theta_{12,p}|$ and 
$\theta_{\star p}+\theta_{12,p}$ (see Eqs.~\ref{eq:tmin}-\ref{eq:tmax}; note that since 
$L_p\gg L_{12}=L_1+L_2$ for this system, $\hat\bJ$ is aligned with $\bl_p$).
The observed stellar obliquity ($\lambda \ge 37^\circ$) then implies
$\theta_{\star p}+\theta_{12,p}\ge 37^\circ$ (see Li et al.~2014 for a previous analysis).

\section{Discussion}

A main goal of this paper is to inform the readers that there is a
simple way to assess the dynamical role of external perturbers on the
spin-orbit misalignments of planetary systems. The key is to evaluate
the star-planet spin-orbit coupling parameter (Eq.~\ref{eq:epss1}) in
response to the differential precession induced by the external
companion. We have presented analytic results in various limiting
regimes, and discussed how the general results (see Fig.~1) scale with
different physical parameters (Section 2). Similar analysis can be
done for multi-planet systems (Section 4), with the generalized
spin-orbit coupling parameter given by Eq.~(\ref{eq:epss12}).

As alluded to in Section 1, continuing observations on spin-orbit
misalignments and external companions in hot Jupiter (HJ) and other compact
planetary systems may shed light on the formation mechanism of
close-in planets. Several recent works of such systems have either
neglected some key physical ingredients (spin-orbit coupling) or made
such a relatively simple problem unnecessarily complex or obscure.
We hope that the results of this paper (see also Lai 2016) 
will make future analysis simpler and more transparent.

A recent analysis on spin-orbit alignments of HJs around cool
stars concluded that the exterior companions are coplanar (Becker et
al.~2017).  While this may be the case of a few systems with gas giant
companion at $\sim 1$~au, similar conclusions cannot be drawn for
systems with a slightly more distant companion (see Section 3,
especially Fig.~\ref{fig2}) since the spin-orbit coupling parameter
$\epsilon_{\star 1}$ depends sensitively on the perturber's semi-major
axis ($\epsilon_{\star 1} \propto a_p^{-3}$; see Eq.~\ref{eq:eq22}).

Certain HJ formation mechanisms (e.g. Lidov-Kozai migration) require
an external companion with a high inclination ($>40^\circ$).  As a
result, constraints placed on the mutual inclination via the method
discussed in this paper may in principle lead to a constraint on the
migration mechanism.  However, we emphasize that the inclination of
giant planet companion may only be constrained when it is relatively
close (at ~1 au, so that $\epsilon_{\star 1}\gtrsim 1$, see
Eq.~\ref{eq:eq22}), such as for WASP-41 and HAT-P-13.  These systems
clearly could not have formed in the classic LK migration picture
(with the HJ originating at ~1 au or beyond), because the initial
configuration would be unstable.  Lidov-Kozai migration requires a
much more distant companion (at, say ~5-10 au, so that the initial
configuration is stable). Such a companion will satisfy
$\epsilon_{\star 1} \ll 1$, so that it is unable to affect the star-HJ
misalignment, and therefore no constraint on the mutual inclination
may be inferred.  We conclude that in practice, using low stellar
obliquities to constrain mutual inclinations is unlikely to be useful
in identifying/ruling out whether Lidov-Kozai migration occurred in HJ
systems.

\section*{Acknowledgments}

This work has been supported in part by NASA grants NNX14AG94G and
NNX14AP31G, and NSF grant AST-1715246. KRA is supported by a NSF graduate
fellowship and BP by a NASA graduate fellowship (NESSF).


\end{document}